\documentclass[]{AO4ELT}  

\usepackage{microtype}
\usepackage[sorting=none]{biblatex}
\usepackage{amsmath,amsfonts,amssymb}
\usepackage{graphicx}
\usepackage{pst-all} 
\usepackage[colorlinks=true, allcolors=blue]{hyperref}
\makeatletter         
\def\@maketitle{
\includegraphics[width = 170mm]{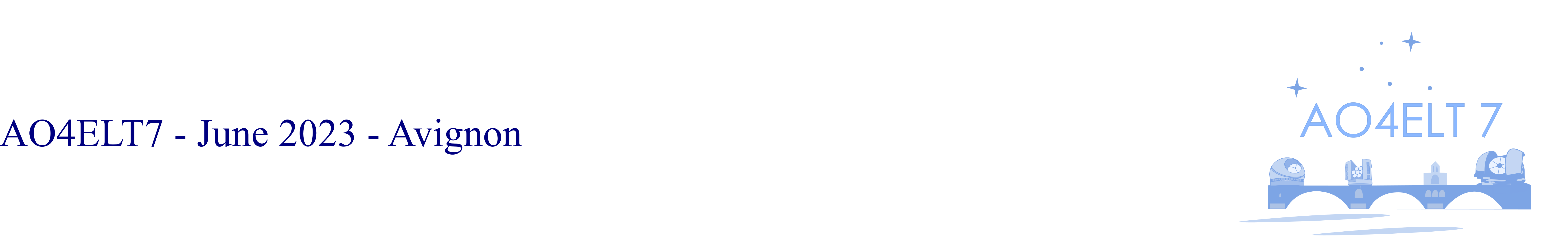}\\[8ex]
\begin{center}
{\Huge \bfseries \sffamily \@title }\\[4ex] 
{\Large  \@author}\\[4ex] 
\@date
\end{center}}

\usepackage[sc, hang]{caption}
\usepackage{graphics}
\usepackage{makeidx}
\usepackage[pass]{geometry}
\usepackage{booktabs}
\usepackage{marvosym}
\usepackage[hang]{subfigure}
\usepackage{rotating}
\usepackage{csquotes}
\usepackage{url}
\usepackage{multicol}
\usepackage{multirow}
\usepackage{overpic}
\usepackage{commath}
\usepackage{bm}
\usepackage{rotating}
\usepackage{psfrag}
\usepackage{parallel}
\usepackage{fancyhdr}

\usepackage[english]{babel} 
\usepackage{eso-pic,graphicx}
\usepackage{enumerate}
\usepackage{psfrag}
\usepackage{booktabs}

\usepackage{color, colortbl}
\definecolor{LightGray}{gray}{0.9}
\usepackage{diagbox}
\usepackage{array}
\newcolumntype{C}[1]{>{\centering\let\newline\\\arraybackslash\hspace{0pt}}m{#1}}

\title{TIPTOP: cone effect for single laser adaptive optics systems}

\author[a,e]{Guido Agapito}
\author[a,e]{Cédric Plantet}
\author[a,e]{Fabio Rossi}
\author[a,e]{Giulia Carlà}
\author[a,e]{Anne-Laure Cheffot}
\author[b,e]{Daniele Vassallo}
\author[c]{Arseniy Kuznetsov}
\author[d]{Simon Conseil}
\author[d]{Benoit Neichel}
\affil[a]{INAF -- Osservatorio Astrofisico di Arcetri, Largo E. Fermi 5, 50125, Firenze, Italy}
\affil[b]{INAF -- Osservatorio Astronomico di Padova, Vicolo dell'Osservatorio, 5, 35122, Padova, Italy}
\affil[c]{European Southern Observatory, Karl-Schwarzschild-str-2, D-85748 Garching, Germany}
\affil[d]{Aix Marseille University, CNRS, CNES, LAM, Marseille, France}
\affil[e]{ADaptive Optics National laboratory in Italy (ADONI)}

\authorinfo{Further author information: (Send correspondence to G.A.)\\G.A.: E-mail: guido.agapito@inaf.it}

\begin{document} 
\maketitle

\begin{abstract}
TIPTOP is a python library that is able to quickly compute Point Spread Functions (PSF) of any kind of Adaptive Optics systems. This library has multiple objectives: support the exposure time calculators of future VLT and ELT instruments, support adaptive optics systems design activities, be part of PSF reconstruction pipelines and support the selection of the best asterism of natural guide stars for observation preparation. Here we report one of the last improvements of TIPTOP: the introduction of the error given by a single conjugated laser, commonly known as the cone effect. The Cone effect was not introduced before because it is challenging due to the non-stationarity of the phase.
Laser guide stars are at a finite distance with respect to the telescope and probe beam accepted by the wavefront sensor has the shape of a cone. Given a single spatial frequency in an atmospheric layer, the cone effect arises from the apparent magnification or stretching of this frequency when it reaches the wavefront sensor. The magnification effect leads to an incorrect estimation of the spatial frequency. Therefore, we estimate the residual power by calculating the difference between two sinusoids with different periods: the nominal one and the magnified one. Replicating this for each spatial frequency we obtain the power spectrum associated with the cone effect. We compare this estimation with the one given by end-to-end simulation and we present how we plan to validate this with on-sky data.
\end{abstract}



\section{Introduction}
\label{sec:intro}  

Adaptive Optics (AO) is a technique used to compensate the effects of atmospheric turbulence and it is becoming increasingly popular among current and future astronomical instruments of large ground based telescopes (a non exahustive list is \cite{2019A&A...631A.155B}, \cite{Macintosh12661}, \cite{2020SPIE11447E..1SC}, \cite{2021arXiv210107091P}, \cite{2021Msngr.182...13C}, \cite{2021Msngr.182...27M}, \cite{2021Msngr.182....7T} and \cite{2022SPIE12185E..08R}).

Numerical simulations are playing an important role in the sizing, development, and analysis of AO systems and several libraries are used to numerically simulate AO systems, like Refs. \cite{2010JEOS....5E0055J}, \cite{2014SPIE.9148E..6CC}, \cite{2016SPIE.9909E..7JC}, \cite{doi:10.1117/12.2233963} and \cite{2016SPIE.9909E..7FR}.
TIPTOP~\cite{2020SPIE11448E..2TN} is one of these libraries: it is an analytical simulator developed in Python that works in the Fourier domain that can quickly\footnote{especially when GPU/CUDA acceleration is available.} produce the AO Point Spread Function (PSF) of several kind of systems.
It is available on github at \url{https://github.com/astro-tiptop/TIPTOP} and on pypi at \url{https://pypi.org/project/astro-tiptop/}.
The goals of TIPTOP are many: from supporting AO system design and observation preparation, estimating the PSF for exposure time calculation, to being part of the PSF reconstruction activity (it is also part of the tools developed in the context of the ESO ELT working groups, see Ref. \cite{2022Msngr.189...23P}).

In this work we focused on a new feature of TIPTOP\footnote{available from version 1.1.}, the ability to simulate the cone effect of a Single Laser guide star Adaptive Optics (SLAO) system like ERIS~\cite{2023A&A...674A.207D} and KECK Laser Guide Star (LGS) AO system~\cite{1998SPIE.3353..260F,2016SPIE.9909E..0SC}.

In the past, attempts have been made to find a solution to introduce the cone effect into Fourier simulation (for example in Ref. \cite{2004SPIE.5490.1356G} and \cite{2008SPIE.7015E..4AA}).
As described in Ref. \cite{neich2008}, in the case of the classical AO, modeling the cone effect using a filter on the PSD is not feasible due to the non-stationarity of the residual phase.
In the case of tomographic AO, the situation is different, as information from several stars is used for volume estimation.
The residual phase is no longer given by the simple difference of a plane wave and a spherical wave, but from the result of turbulent volume estimation by several spherical waves and the plane wave.
The residual phase thus regains a certain stationary character.

In this work we followed a different approach based on sinusoids and on the magnification effect seen by the wavefront sensor and given by the spherical wave propagation.
This method is described in Sec.~\ref{sec:coneFreq} and it is combined with the other error sources of TIPTOP and the tilt filter (tilt filter is presented in Sec.~\ref{sec:tilt_filt}) to produce the so called high orders PSF, that is the PSF generated by the LGS correction without the tip and tilt errors associated with the sensing and correction of the natural guide star.
We compared the results from TIPTOP with the one of end-to-end simulation in Sec.~\ref{sec:sim}.
Finally, we conclude the paper with a few comments in Sec.~\ref{sec:concl} and we present a possible alternative approach in Appendix~\ref{sec:altern}.

\section{Cone effect in the spatial frequency domain}\label{sec:coneFreq}

%
\begin{figure}[h]
    \centering
    \subfigure[Scheme of the geometry. Black curve is the input signal on the right.\label{fig:ConeEffectSF}]{
    \includegraphics[width=0.3\linewidth]{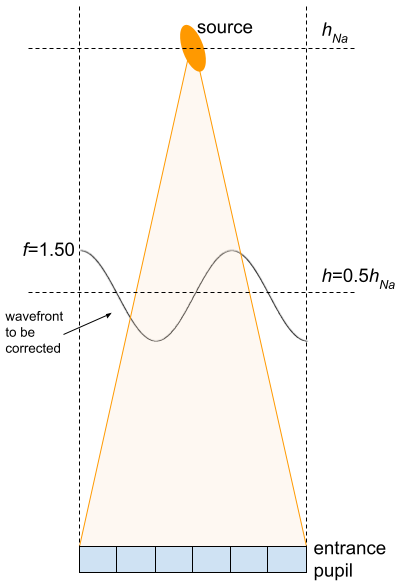}}
    \hspace{5mm}
    \subfigure[Input signal (wavefront to be corrected), reconstructed signal (wavefront seen) and error, 0 phase shift case. X-axis units are radius units.\label{fig:signalExample}]{
    \includegraphics[width=0.55\linewidth]{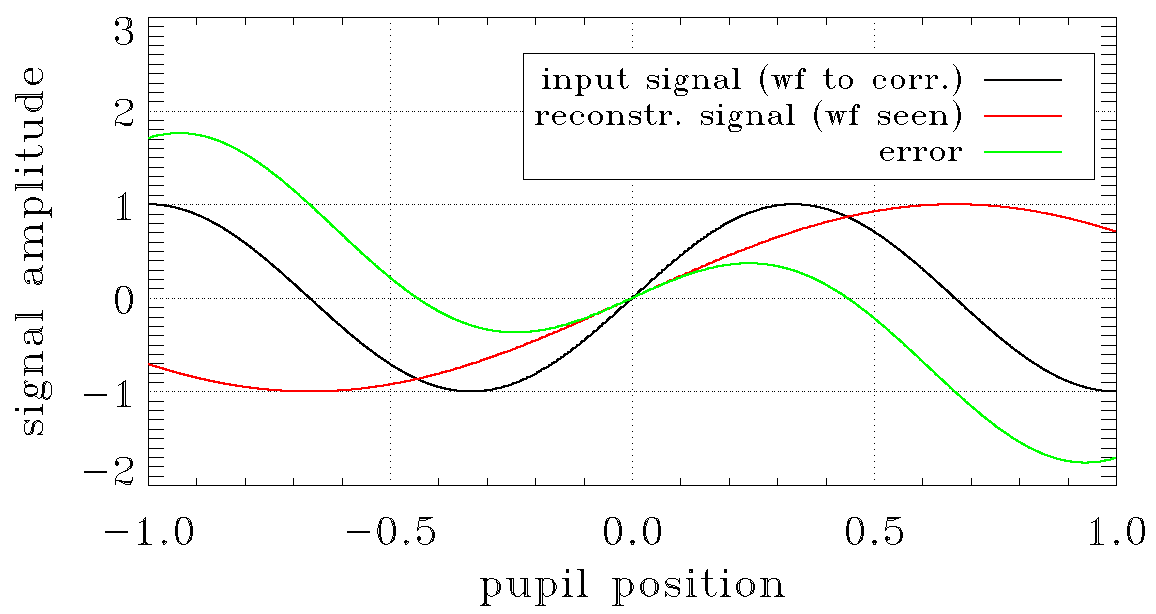}}
    \caption{Example of the magnification given by the cone effect on a spatial frequency with 1.5 period in the diameter at a distance from the entrance pupil that is half the distance between the source and the entrance pupil. The sensed signal has a spatial frequency that is half the one of the input.}
\end{figure}
%
%
%
\begin{figure}[h]
    \centering
    \includegraphics[width=0.65\linewidth]{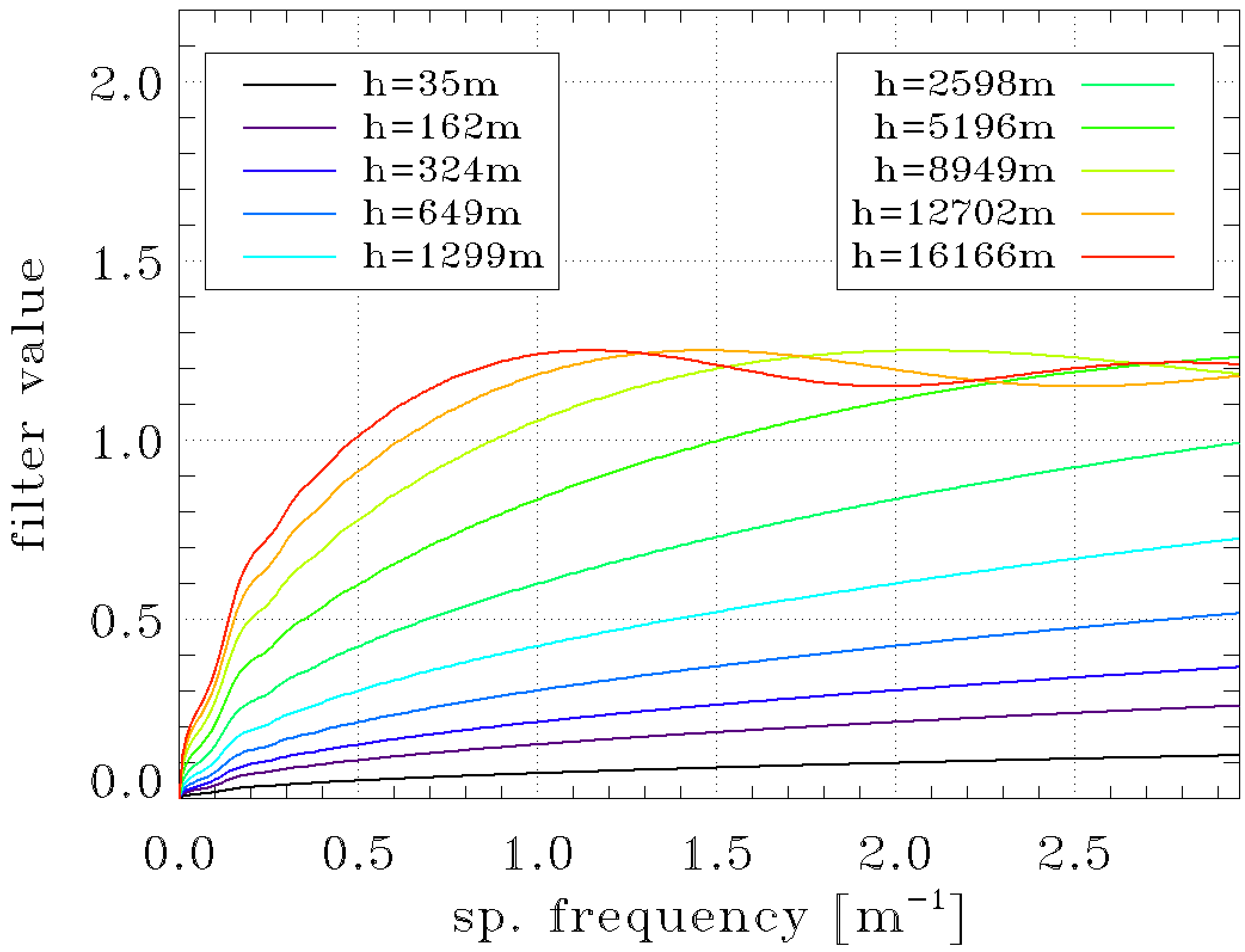}
    \caption{Single dimension filter values from the brute force approach for an 8 m telescope with source at 90 km, zenith angle of 30 deg and a $Cn^2$ profile with altitudes of $[30, 140, 281, 562, 1125, 2250, 4500, 7750, 11000, 14000]$ m.}
    \label{fig:filterSingleDim}
\end{figure}
\begin{figure}[h]
    \centering
    \includegraphics[width=0.95\linewidth]{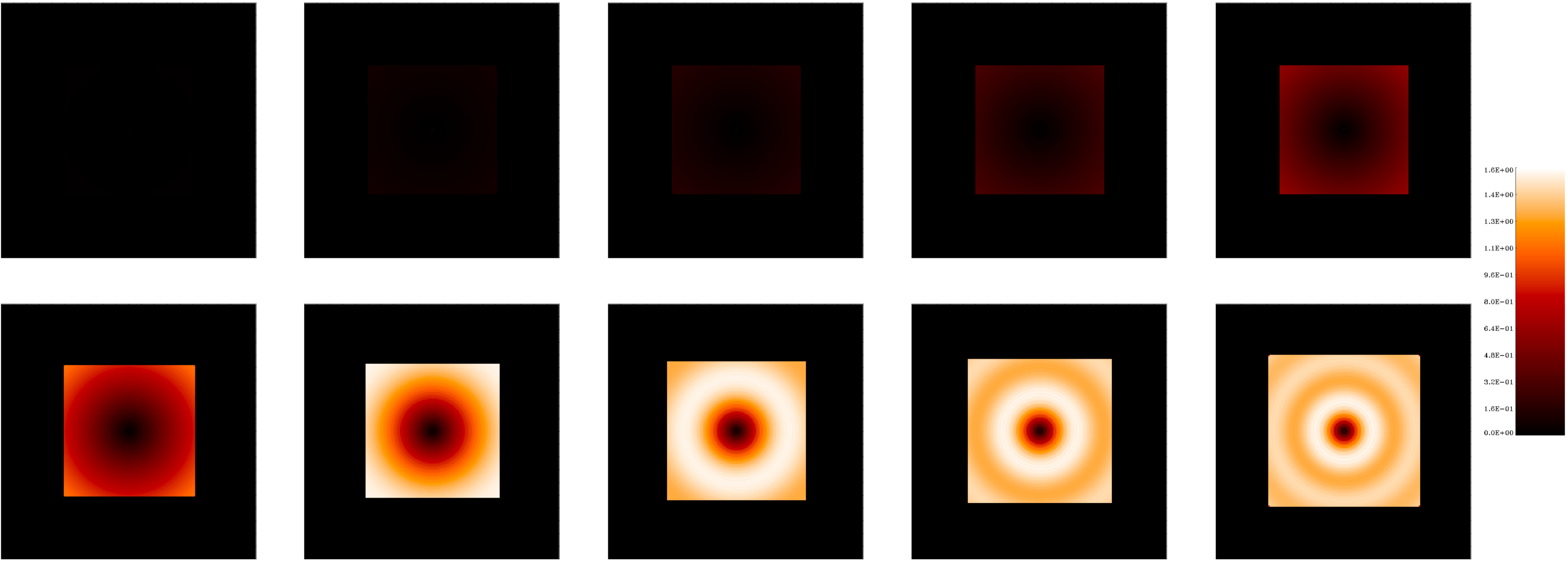}
    \caption{2D filter values from the brute force approach for an 8 m telescope with source at 90 km, zenith angle of 30 deg and a $Cn^2$ profile with altitudes of $[30, 140, 281, 562, 1125, 2250, 4500, 7750, 11000, 14000]$ m (the first 5 values correspond to the first row of images, the last 5 values to the second row). Spatial frequencies range from -5 to 5 m$^{-1}$, with a cut at 2.5 m$^{-1}$ due to the DM pitch of 0.2 m.}
    \label{fig:filterCone}
\end{figure}
\begin{figure}[h]
    \centering
    \includegraphics[width=0.75\linewidth]{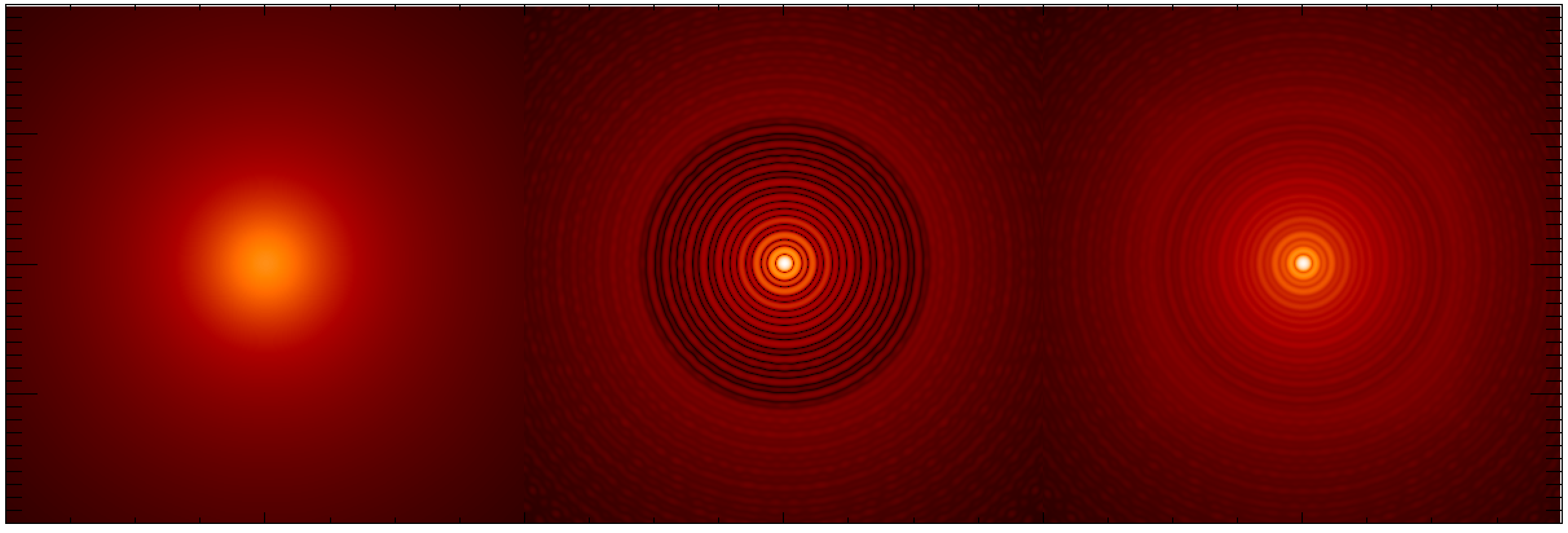}
    \caption{Comparison of estimated PSFs considering a $r_0$(500 nm)=0.14 m on the line of sight. Left, seeing limited, center, fitting error only (pitch=0.2m), right, fitting error and cone effect. The SR in K band (2200nm) is 0.92 in case of fitting error only and 0.76 in case of fitting error and cone effect for a source at 90 km (based on the brute force computations).}
    \label{fig:PSFcomparison}
\end{figure}

In this section we present the method to compute the error associated with the so called focal anisoplanatism or cone effect for a SLAO system.
Let's consider a single spatial frequency $f<f^{DM}_c$ (DM is deformable mirror, and $f^{DM}_c$ the DM's maximum correction frequency linked to the actuator pitch).
The correction of $f$ in a system with the source at infinity is perfect, if we exclude aliasing, measurement errors, and reconstruction errors. However, in a system with a source at finite distance, the wavefront sensor can be tricked by a magnification effect due to the systems geometry. As shown in Fig. \ref{fig:ConeEffectSF}), only the central part of the wavefront is measured by the sensor.
This measurement is then used to control the DM on the full pupil.
This leads to a mismatch between the wavefront seen by the observed star and the corrected wavefront.
The magnification of the wavefront at a distance $h$ from the entrance pupil is equal to
\begin{equation}
    \label{eq:Magn}
    m=\frac{h_{Na}}{h_{Na}-h} \; .
\end{equation}
With $h_{Na}$ the altitude of the LGS and $h$ the altitude of the atmospheric layer observed.
If $x$ is the pupil position going from $-r$ to $r$ ($r$ is pupil radius) and considering a sinusoidal signal with unitary amplitude, $s(f,x)=\sin(2\pi fx)$, the reconstructed signal, $s_{rec}$, is:
\begin{equation}
    \label{eq:sinRec}
    s_{rec}(f,x,m,\phi)=A(f,m)\sin \left( 2\pi \frac{f}{m} x+\phi \right)
\end{equation}
where $\phi$ is the phase shift and $A(f,m)$ is approximated as:
\begin{equation}
    \label{eq:amp}
    A(f,m,\phi) = \arg \min_A \left( \sqrt{\int_{-r}^{r} \left( \sin(2\pi fx+\phi) - A\sin \left( 2\pi \frac{f}{m} x+\phi \right) \right) x^2 dx }\right) \; .
\end{equation}
The minimization found in equation \ref{eq:amp} is an approximation of the closed loop behaviour.
The AO closed loop is able to minimize the measurement of the phase seen by the WFS but on $x=[-r/m,r/m]$ instead of $x=[-r,r]$.
Additionally there are some cross-talks between spatial frequencies.
Note that in open loop A is 1 if $f/m>0.25/r$ and $A=\frac{\sin \left( 2\pi \frac{f}{m} r \right)}{\sin(2\pi f r)}$ otherwise.

The error, $e(f,x,m)$, is:
\begin{equation}
    \label{eq:err}
    e(f,m,x,\phi) = s(f,x,\phi) - s_{rec}(f,x,m,\phi) = \sin(2\pi fx+\phi) - A(f,m,\phi)\sin \left( 2\pi \frac{f}{m}x+\phi \right)
\end{equation}
An example of this error for $\phi$=0 is shown in Fig. \ref{fig:signalExample}, and this error can be used to derive the filtering function.
We approximate that the spatial frequencies found in the error signal are the same as the one seen at the wavefront sensor.
This approximation is negligible when the magnification is close to 1, but can introduce large errors when the magnification is much greater than 1.
Fortunately, the maximum value of $m$ is around 1.3 when considering sodium laser ($h_{Na}\simeq$ 90 km) and standard $C_n^2$ profile ($\mathrm{max}(h)\simeq$ 20 km).
The filter coefficient, $k(f,m)$, is described as:
\begin{equation}
    \label{eq:filter}
    k(f,m) = \sum_{j=0}^{N_{\phi}-1}\frac{1}{N_{\phi}} \frac{\sigma_e(f,m,\frac{2\pi i}{N_{\phi}} )}{\sigma_s \left( f,m,\frac{2\pi i}{N_{\phi}} \right)}
\end{equation}
where $\sigma_e$ is the standard deviation of $e$, the error, $\sigma_s$ is the standard deviation of $s$, the signal and $N_{\phi}$ is the phase shift.
An example of the filter coefficients, $k$, for an 8m telescope is shown in Fig. \ref{fig:filterSingleDim}.

As depicted in Figure \ref{fig:filterCone}, the cone effect in the spatial frequency domain has circular symmetry and depends only on the norm of the frequency vector $\mathbf{f}=(f_x,f_y)$. This simplifies the computation of $k(f_x,f_y)$.
Using a PSD the cone effect can be described as:
\begin{equation}
    \label{eq:psdConeBrute}
    W_{cone}(\mathbf{f}) = \sum_{i=1}^{N_{layer}}C_n^2(i)k^2(\mathbf{f},m_{i}) W_{turb}(\mathbf{f})
\end{equation}
where $0<C_n^2(i)\leq1$ is the fraction of $C_n^2$ associated to the layer $i$ and $W_{turb}$ is the PSD of the turbulence.
An example of the PSFs produced with this estimation is reported in Fig. \ref{fig:PSFcomparison}.

\section{Tilt filter}\label{sec:tilt_filt}
\begin{figure}[h]
    \centering
    \subfigure[]{
    \includegraphics[width=0.45\linewidth]{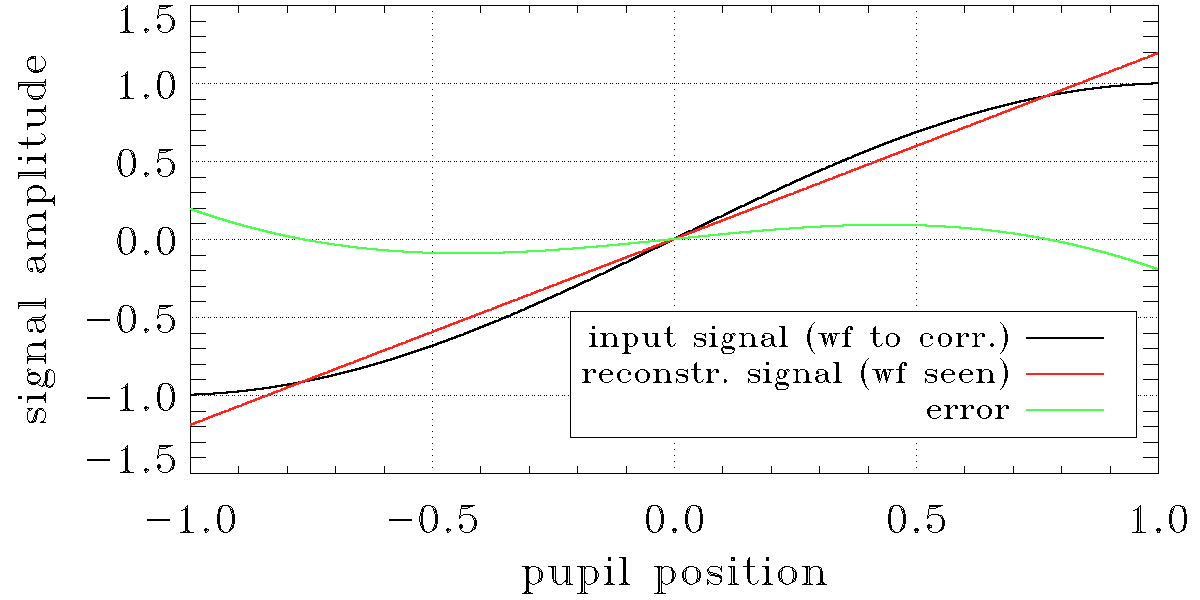}}
    \hspace{5mm}
    \subfigure[]{
    \includegraphics[width=0.45\linewidth]{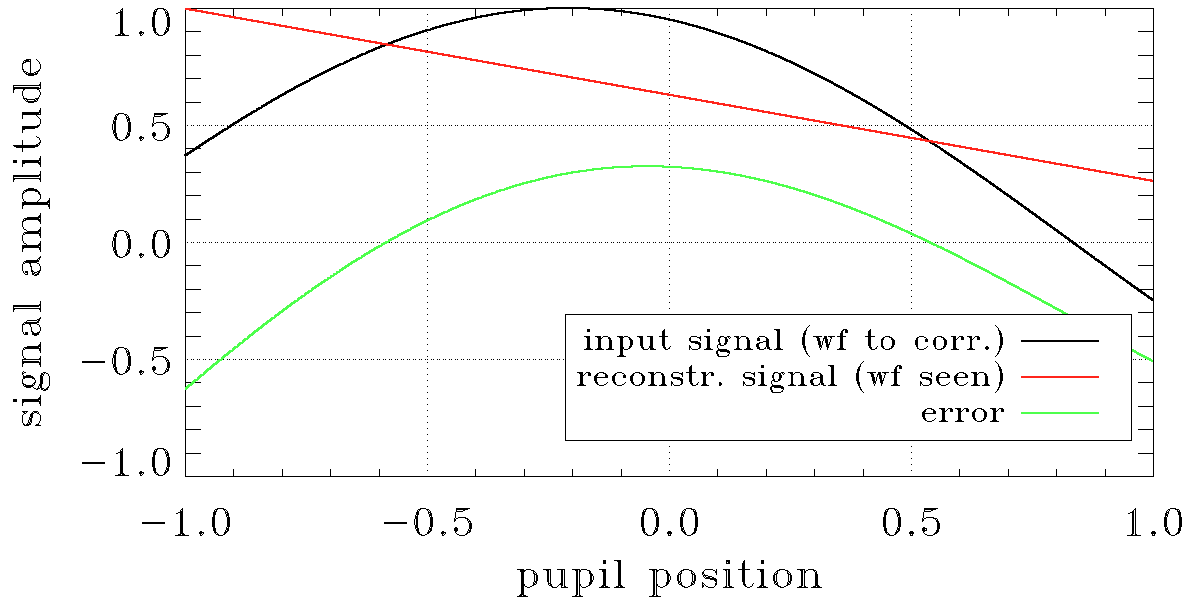}}
    \caption{Example of the tilt estimated on a sinusoidal with a period twice the pupil size. (a), 0deg phase shift, (b), 108deg of phase shift.}
    \label{fig:signalExample_tilt}
\end{figure}
\begin{figure}[h]
    \centering
    \includegraphics[width=0.65\linewidth]{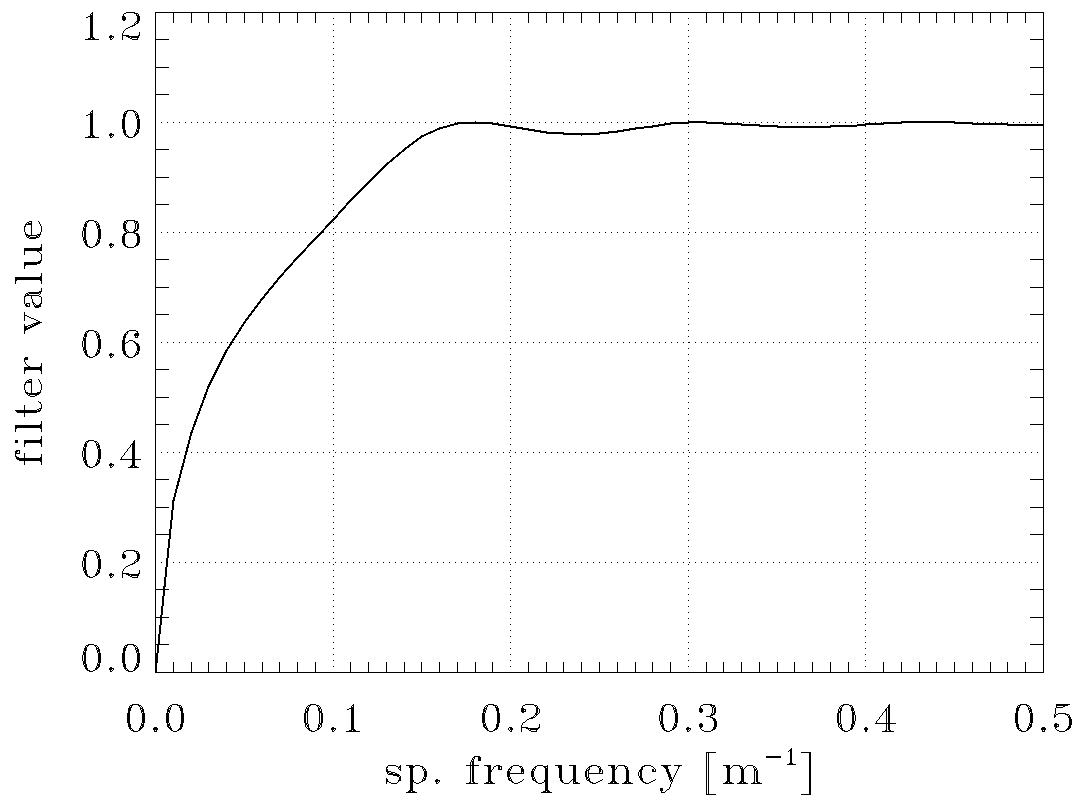}
    \caption{Single dimension tilt filter values.}
    \label{fig:filterSingleDim_tilt}
\end{figure}
\begin{figure}[h]
    \centering
    \includegraphics[width=0.75\linewidth]{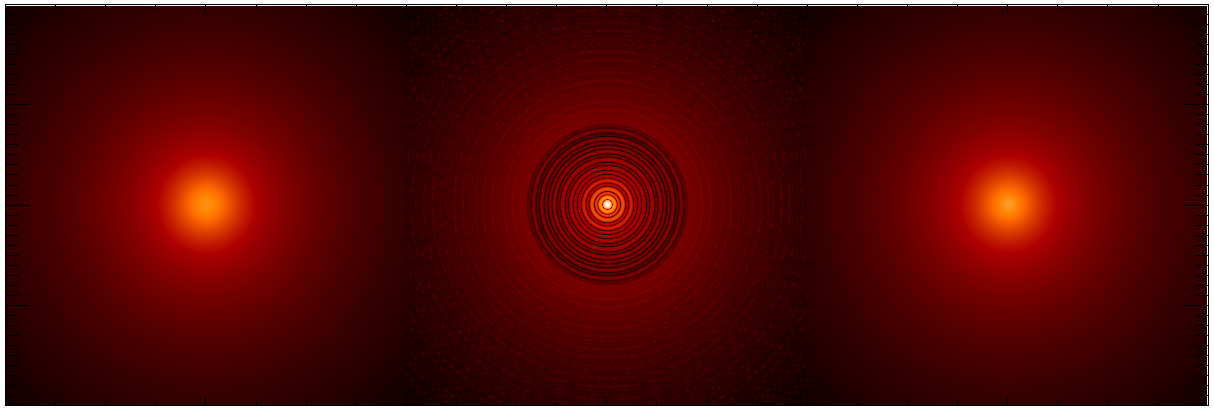}
    \caption{Comparison of estimated PSFs considering a $r_0$(500nm)=0.14m on the line of sight. Left, seeing limited, center, fitting error only (pitch=0.2m), right seeing limited without tilt. SR in K band (2200nm) is 0.92 in case of fitting error only and 0.03 in case of seeing limited without tilt.}
    \label{fig:PSFcomparison_tilt}
\end{figure}
As an approximation in laser-based AO systems, the computation of the high-order PSF was performed using only a piston filter in TIPTOP.
The calculation of the high order PSF should exclude any errors related to the tilt as they have already been accounted for in the low-order calculation.
To achieve this, a tilt filter is required for the high order PSD.
In this section, we present this tilt filter and follow the same method as the one in Section~\ref{sec:coneFreq} for the cone effect.
The base idea is to compute what should be the tilt contribution to the high order depending on the atmospheric layer we are looking at, and use that information to remove the tilt from the PSD.

Let's consider a single spatial frequency $f<f^{DM}_c$, impacted by a tilt offset.
The tilt filter coefficient for this frequency will be given by:
\begin{equation}
    \label{eq:tiltFilterCoeff}
    k(f) = \sum_{j=0}^{N_{\phi}-1}\frac{1}{N_{\phi}}\frac{\sigma_e(f,2\pi i/N_{\phi})}{\sigma_s(f,2\pi i/N_{\phi})} 
\end{equation}
where $\sigma$ is the standard deviation, $e(f,x,\phi) = s(f,x,\phi) - s_{lin}(f,x,\phi)$, $s(f,x,\phi)=\sin(2\pi f x+\phi)$ and $s_{lin}$ is the linear fitting of $s$.
An example of these signals is shown in Fig. \ref{fig:signalExample_tilt}, and the filter coefficients, $k$, for an 8m telescope is show in Fig. \ref{fig:filterSingleDim_tilt}.
As for the cone effect, the tilt filter in the spatial frequency plane has a circular symmetry, so we can compute the filter map, $k(f_x,f_y)$, deduce a PSD from it, and remove that PSD from the high-order PSD.
An example of the PSFs produced with this estimation is reported in Fig. \ref{fig:PSFcomparison_tilt}.

Finally, we compared TIPTOP with the end-to-end simulator PASSATA~\cite{doi:10.1117/12.2233963}.
We compute the amount of power associated with tip and tilt in the case of a line-of-sight seeing of 0.87 arcsec.
During the 5-second simulation performed using PASSATA, it was observed that the quadratic sum of the two tilt axis exhibits 900nm Root Mean Square (RMS).
By comparison, using TIPTOP resulted in an 850nm RMS.
TIPTOP's value is lower, yet the agreement remains satisfactory and suitable for a single occurrence of relatively brief turbulence in PASSATA.
Additionally, the total turbulence RMS of TIPTOP for this situation is 1240 nm and of PASSATA 1230 nm.

\section{Results}\label{sec:sim}
\begin{figure}[h]
    \centering
    
    \subfigure[LGS source altitude is 90km (z=0deg). The ratio of the K band SR (TIPTOP over PASSATA) is 0.97.\label{fig:TIPTOPandPASSATA}]{
    \includegraphics[width=0.85\linewidth]{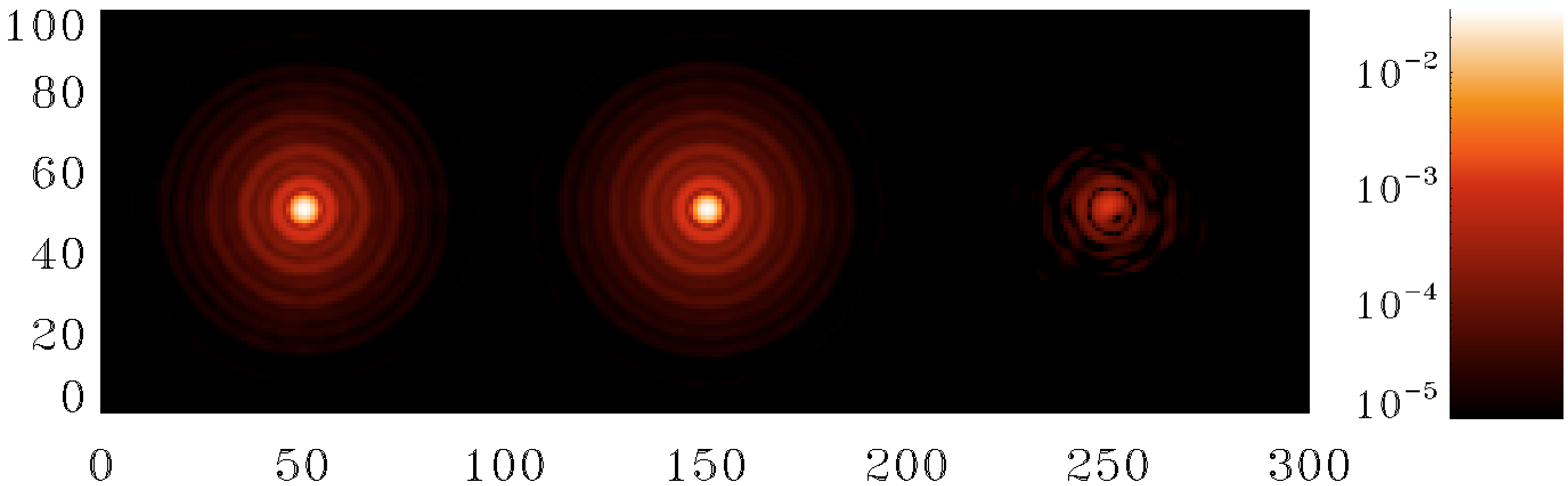}}

    \subfigure[LGS source altitude is 45km (z=0deg). The ratio of the K band SR (TIPTOP over PASSATA) is 0.93.\label{fig:TIPTOPandPASSATA45km}]{
    \includegraphics[width=0.85\linewidth]{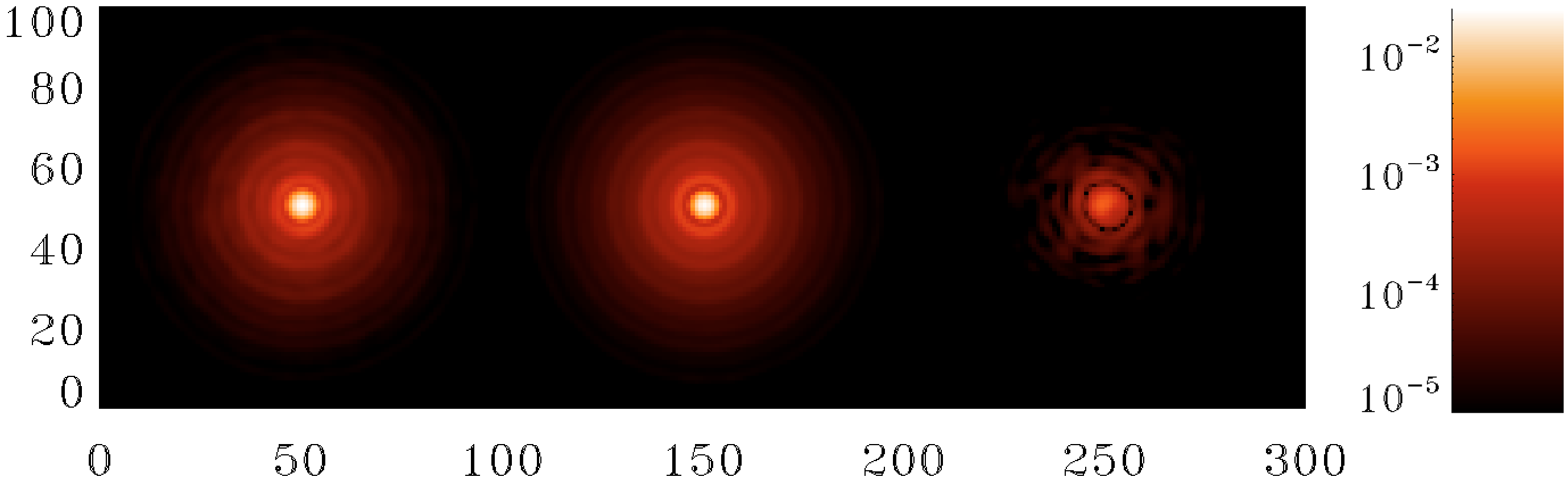}}
    
    \caption{Comparison of estimated PSFs in K band (2200nm) for the ERIS LGS mode. Left PASSATA, center TIPTOP and right absolute value of the difference.}
    \label{fig:TIPTOPandPASSATAall}
\end{figure}
In this section we present the results of a simulation of a single LGS AO system like ERIS~\cite{2023A&A...674A.207D}.
We compared TIPTOP with an end-to-end simulator, PASSATA~\cite{doi:10.1117/12.2233963}.
We consider an atmosphere with a seeing of 0.87 arcsec at the zenith and an average wind speed of ~11m/s.
The laser sensing is given by a 40$\times$40 Shack-Hartmann sensor while the visible WFS of the natural guide star is a 4$\times$4 Shack-Hartmann sensor.
The correction is given by a deformable mirror with 1156 actuators~\cite{2017Msngr.168....8A}.
The comparison is shown in Fig. \ref{fig:TIPTOPandPASSATA} considering a zenith angle of 0 deg and a laser spot at 90km.
An additional test with a LGS source at 45km is reported in Fig. \ref{fig:TIPTOPandPASSATA45km}.
As can be seen in these two figures the agreement of TIPTOP with an end-to-end simulation is good with a difference of a few percents.

\section{Conclusion}\label{sec:concl}

We presented a new feature of TIPTOP, the ability to simulate the error associated to the so called focal anisoplanatism, or cone effect.
This feature enables TIPTOP to estimate the PSF of AO systems like ERIS, KECK LGS AO system or the future GRAVITY+ systems~\cite{2022Msngr.189...17A}.
We show that this new feature gives results consistent with the end-to end simulation PASSATA~\cite{doi:10.1117/12.2233963} of the aforementioned systems. 
TIPTOP is a python library under development. We heavily rely on user feedback to improve the library and provide new functionality. 
The aim for the future of TIPTOP is to provide a library that is both stable and accurate, with ample examples and improved documentation that can be utilised by the 
entire AO community.

\begin{appendix}
\section{A possible alternative approach}\label{sec:altern}

%
%
%
In this section we present a possible alternative approach for computing the cone effect error (in addition to the one presented in Ref. \cite{2004SPIE.5490.1356G}  and \cite{2008SPIE.7015E..4AA}) in TIPTOP.
This approach is based on the statistics of the turbulent phase: starting from eq. 24. of Ref. \cite{2022JOSAA..39...17P} and considering that $R=R_1=R_2$, $A_{1l}=0$ and $A=A_{2l}$ we can write:
\begin{equation}
    \label{eq:PSDcone1}
    W_{cone}^i(\mathbf{f}) = W_{\phi_l}(\mathbf{f}) \Bigg[ \frac{J_1 \left(2\pi R (K-1) f \right)}{\pi R (K-1) f} - \frac{J_1 \left(2\pi R f \right)}{\pi R f} \frac{J_1 \left(2\pi R (1-A) f \right)}{\pi R (1-A) f} \Bigg]
\end{equation}
where $W_{cone}^i(\mathbf{f})$ is the PSD related to cone effect for the layer $i$. Then if we make the difference with the classical filter function for piston removal we get:
\begin{equation}
    \label{eq:PSDcone2}
    W_{cone}^i(\mathbf{f}) = W_{\phi_l}(\mathbf{f}) \Bigg[ 1-\left( \frac{J_1 \left(2\pi R f \right)}{\pi R f} \right)^2 - \frac{J_1 \left(2\pi R (K-1) f \right)}{\pi R (K-1) f} + \frac{J_1 \left(2\pi R f \right)}{\pi R f} \frac{J_1 \left(2\pi R (1-A) f \right)}{\pi R (1-A) f} \Bigg]
\end{equation}
that can be written as:
\begin{equation}
    \label{eq:PSDcone3}
    W_{cone}^i(\mathbf{f}) = W_{\phi_l}(\mathbf{f}) \Bigg[ 1- \frac{J_1 \left(2\pi R A f \right)}{\pi R A f} +
    \frac{J_1 \left(2\pi R f \right)}{\pi R f} \left( \frac{J_1 \left(2\pi R (1-A) f \right)}{\pi R (1-A) f} - \frac{J_1 \left(2\pi R f \right)}{\pi R f} \right)
    \Bigg]
\end{equation}
This equation gives the power spectral density (PSD) of the error associated with the cone effect.
Please note that:
\begin{itemize}
    \item $A$ is the ratio $h/h_{Na}$, where $h_{Na}$ is the distance between the pupil to the laser guide star and $h$ is the distance between the the pupil and the observed atmospheric layer. A is therefore the description of how the atmospheric layer is stretched because of the cone effect.
    \item $K=1-A$
\end{itemize}
%
\end{appendix}



\printbibliography 

@INPROCEEDINGS{doi:10.1117/12.2233963,
       author = {{Agapito}, G. and {Puglisi}, A. and {Esposito}, S.},
        title = "{PASSATA: object oriented numerical simulation software for adaptive optics}",
    booktitle = {Adaptive Optics Systems V},
         year = 2016,
       editor = {{Marchetti}, Enrico and {Close}, Laird M. and {V{\'e}ran}, Jean-Pierre},
       series = {Society of Photo-Optical Instrumentation Engineers (SPIE) Conference Series},
       volume = {9909},
        month = jul,
          eid = {99097E},
        pages = {99097E},
          doi = {10.1117/12.2233963},
archivePrefix = {arXiv},
       eprint = {1607.07624},
 primaryClass = {astro-ph.IM},
       adsurl = {https://ui.adsabs.harvard.edu/abs/2016SPIE.9909E..7EA}
}

@INPROCEEDINGS{2022SPIE12185E..08R,
       author = {{Riccardi}, A. and {Puglisi}, A. and {Grani}, P. and {Briguglio}, R. and {Esposito}, S. and {Agapito}, G. and {Biliotti}, V. and {Bonaglia}, M. and {Carbonaro}, L. and {Xompero}, M. and {Baruffolo}, A. and {Salasnich}, B. and {Di Rico}, G. and {Davies}, R. and {Feuchtgruber}, H. and {Rau}, C. and {Dallilar}, Y. and {Kravchenko}, K. and {Kolb}, J. and {Haguenauer}, P. and {Soenke}, C. and {Barr}, D. and {Cortes}, A. and {Reyes}, J.},
        title = "{The ERIS Adaptive Optics System: first on-sky results of the ongoing commissioning at the VLT-UT4}",
    booktitle = {Adaptive Optics Systems VIII},
         year = 2022,
       editor = {{Schreiber}, Laura and {Schmidt}, Dirk and {Vernet}, Elise},
       series = {Society of Photo-Optical Instrumentation Engineers (SPIE) Conference Series},
       volume = {12185},
        month = aug,
          eid = {1218508},
        pages = {1218508},
          doi = {10.1117/12.2629425},
       adsurl = {https://ui.adsabs.harvard.edu/abs/2022SPIE12185E..08R}
}

@ARTICLE{2021Msngr.182...13C,
       author = {{Ciliegi}, P. and {Agapito}, G. and {Aliverti}, M. and {Annibali}, F. and {Arcidiacono}, C. and {Balestra}, A. and {Baruffolo}, A. and {Bergomi}, M. and {Bianco}, A. and {Bonaglia}, M. and {Busoni}, L. and {Cantiello}, M. and {Cascone}, E. and {Chauvin}, G. and {Chinellato}, S. and {Cianniello}, V. and {Correia}, J. -J. and {Cosentino}, G. and {Dall'Ora}, M. and {De Caprio}, V. and {Devaney}, N. and {Di Antonio}, I. and {Di Cianno}, A. and {Di Giammatteo}, U. and {D'Orazi}, V. and {Di Rico}, G. and {Dolci}, M. and {Dout{\`e}}, S. and {Eredia}, C. and {Farinato}, J. and {Esposito}, S. and {Fantinel}, D. and {Feautrier}, P. and {Foppiani}, I. and {Giro}, E. and {Gluck}, L. and {Golden}, A. and {Goncharov}, A. and {Grani}, P. and {Gullieuszik}, M. and {Haguenauer}, P. and {H{\'e}nault}, F. and {Hubert}, Z. and {Le Louran}, M. and {Magrin}, D. and {Maiorano}, E. and {Mannucci}, F. and {Malone}, D. and {Marafatto}, L. and {Moraux}, E. and {Munari}, M. and {Oberti}, S. and {Pariani}, G. and {Pettazzi}, L. and {Plantet}, C. and {Podio}, L. and {Portaluri}, E. and {Puglisi}, A. and {Ragazzoni}, R. and {Rakich}, A. and {Rabou}, P. and {Redaelli}, E. and {Redman}, M. and {Riva}, M. and {Rochat}, S. and {Rodeghiero}, G. and {Salasnich}, B. and {Saracco}, P. and {Sordo}, R. and {Spavone}, M. and {Sztefek}, M. -H. and {Valentini}, A. and {Vanzella}, E. and {Verinaud}, C. and {Xompero}, M. and {Zaggia}, S.},
        title = "{MAORY: A Multi-conjugate Adaptive Optics RelaY for ELT}",
      journal = {The Messenger},
         year = 2021,
        month = mar,
       volume = {182},
        pages = {13-16},
          doi = {10.18727/0722-6691/5216},
archivePrefix = {arXiv},
       eprint = {2103.11219},
 primaryClass = {astro-ph.IM},
       adsurl = {https://ui.adsabs.harvard.edu/abs/2021Msngr.182...13C}
}

@ARTICLE{2021Msngr.182....7T,
       author = {{Thatte}, N. and {Tecza}, M. and {Schnetler}, H. and {Neichel}, B. and {Melotte}, D. and {Fusco}, T. and {Ferraro-Wood}, V. and {Clarke}, F. and {Bryson}, I. and {O'Brien}, K. and {Mateo}, M. and {Garcia Lorenzo}, B. and {Evans}, C. and {Bouch{\'e}}, N. and {Arribas}, S. and {HARMONI Consortium}},
        title = "{HARMONI: the ELT's First-Light Near-infrared and Visible Integral Field Spectrograph}",
      journal = {The Messenger},
         year = 2021,
        month = mar,
       volume = {182},
        pages = {7-12},
          doi = {10.18727/0722-6691/5215},
archivePrefix = {arXiv},
       eprint = {2103.11215},
 primaryClass = {astro-ph.IM},
       adsurl = {https://ui.adsabs.harvard.edu/abs/2021Msngr.182....7T}
}

@ARTICLE{2022JOSAA..39...17P,
       author = {{Plantet}, C{\'e}dric and {Carl{\`a}}, Giulia and {Agapito}, Guido and {Busoni}, Lorenzo},
        title = "{Spatiotemporal statistics of the turbulent piston-removed phase and Zernike coefficients for two distinct beams}",
      journal = {Journal of the Optical Society of America A},
         year = 2022,
        month = jan,
       volume = {39},
       number = {1},
        pages = {17},
          doi = {10.1364/JOSAA.431520},
archivePrefix = {arXiv},
       eprint = {2209.00931},
 primaryClass = {astro-ph.IM},
       adsurl = {https://ui.adsabs.harvard.edu/abs/2022JOSAA..39...17P}
}

@INPROCEEDINGS{2020SPIE11448E..2TN,
       author = {{Neichel}, Benoit and {Beltramo-Martin}, Olivier and {Plantet}, C{\'e}dric and {Rossi}, Fabio and {Agapito}, Guido and {Fusco}, Thierry and {Carolo}, Elena and {Carl{\`a}}, Giulia and {Cirasuolo}, Michele and {Van Der Burg}, Remco},
        title = "{TIPTOP: a new tool to efficiently predict your favorite AO PSF}",
    booktitle = {Adaptive Optics Systems VII},
         year = 2020,
       editor = {{Schreiber}, Laura and {Schmidt}, Dirk and {Vernet}, Elise},
       series = {Society of Photo-Optical Instrumentation Engineers (SPIE) Conference Series},
       volume = {11448},
        month = dec,
          eid = {114482T},
        pages = {114482T},
          doi = {10.1117/12.2561533},
archivePrefix = {arXiv},
       eprint = {2101.06486},
 primaryClass = {astro-ph.IM},
       adsurl = {https://ui.adsabs.harvard.edu/abs/2020SPIE11448E..2TN}
}

@PHDTHESIS{neich2008,
url = "http://www.theses.fr/2008PA077206",
title = "Etude des galaxies lointaines et optiques adaptatives tomographiques pour ELTs",
author = "Neichel, Benoît",
year = "2008",
pages = "1 vol. (331 p.)",
note = "Thèse de doctorat dirigée par Hammer, François Astrophysique et méthodes associées Paris 7 2008",
note = "2008PA077206",
}

@ARTICLE{2021arXiv210107091P,
       author = {{Pinna}, Enrico and {Rossi}, Fabio and {Puglisi}, Alfio and {Agapito}, Guido and {Bonaglia}, Marco and {Plantet}, Cedric and {Mazzoni}, Tommaso and {Briguglio}, Runa and {Carbonaro}, Luca and {Xompero}, Marco and {Grani}, Paolo and {Riccardi}, Armando and {Esposito}, Simone and {Hinz}, Phil and {Vaz}, Amali and {Ertel}, Steve and {Montoya}, Oscar M. and {Durney}, Oliver and {Christou}, Julian and {Miller}, Doug L. and {Taylor}, Greg and {Cavallaro}, Alessandro and {Lefebvre}, Michael},
        title = "{Bringing SOUL on sky}",
      journal = {arXiv e-prints},
         year = 2021,
        month = jan,
          eid = {arXiv:2101.07091},
        pages = {arXiv:2101.07091},
          doi = {10.48550/arXiv.2101.07091},
archivePrefix = {arXiv},
       eprint = {2101.07091},
 primaryClass = {astro-ph.IM},
       adsurl = {https://ui.adsabs.harvard.edu/abs/2021arXiv210107091P}
}

@ARTICLE{2021Msngr.182...27M,
       author = {{Marconi}, A. and {Abreu}, M. and {Adibekyan}, V. and {Aliverti}, M. and {Allende Prieto}, C. and {Amado}, P. and {Amate}, M. and {Artigau}, E. and {Augusto}, S. and {Barros}, S. and {Becerril}, S. and {Benneke}, B. and {Bergin}, E. and {Berio}, P. and {Bezawada}, N. and {Boisse}, I. and {Bonfils}, X. and {Bouchy}, F. and {Broeg}, C. and {Cabral}, A. and {Calvo-Ortega}, R. and {Canto Martins}, B.~L. and {Chazelas}, B. and {Chiavassa}, A. and {Christensen}, L. and {Cirami}, R. and {Coretti}, I. and {Covino}, S. and {Cresci}, G. and {Cristiani}, S. and {Cunha Parro}, V. and {Cupani}, G. and {de Castro Le{\~a}o}, I. and {Renan de Medeiros}, J. and {Furlande Souza}, M.~A. and {Di Marcantonio}, P. and {Di Varano}, I. and {D'Odorico}, V. and {Doyon}, R. and {Drass}, H. and {Figueira}, P. and {Belen Fragoso}, A. and {Uldall Fynbo}, J.~P. and {Gallo}, E. and {Genoni}, M. and {Gonz{\'a}lez Hern{\'a}ndez}, J. and {Haehnelt}, M. and {Hlavacek-Larrondo}, J. and {Hughes}, I. and {Huke}, P. and {Humphrey}, A. and {Kjeldsen}, H. and {Korn}, A. and {Kouach}, D. and {Landoni}, M. and {Liske}, J. and {Lovis}, C. and {Lunney}, D. and {Maiolino}, R. and {Malo}, L. and {Marquart}, T. and {Martins}, C. and {Mason}, E. and {Molaro}, P. and {Monnier}, J. and {Monteiro}, M. and {Mordasini}, C. and {Morris}, T. and {Mucciarelli}, A. and {Murray}, G. and {Niedzielski}, A. and {Nunes}, N. and {Oliva}, E. and {Origlia}, L. and {Pall{\'e}}, E. and {Pariani}, G. and {Parr-Burman}, P. and {Pe{\~n}ate}, J. and {Pepe}, F. and {Pinna}, E. and {Piskunov}, N. and {Rasilla Pi{\~n}eiro}, J.~L. and {Rebolo}, R. and {Rees}, P. and {Reiners}, A. and {Riva}, M. and {Romano}, D. and {Rousseau}, S. and {Sanna}, N. and {Santos}, N. and {Sarajlic}, M. and {Shen}, T. -C. and {Sortino}, F. and {Sosnowska}, D. and {Sousa}, S. and {Stempels}, E. and {Strassmeier}, K. and {Tenegi}, F. and {Tozzi}, A. and {Udry}, S. and {Valenziano}, L. and {Vanzi}, L. and {Weber}, M. and {Woche}, M. and {Xompero}, M. and {Zackrisson}, E. and {Zapatero Osorio}, M.~R.},
        title = "{HIRES, the High-resolution Spectrograph for the ELT}",
      journal = {The Messenger},
         year = 2021,
        month = mar,
       volume = {182},
        pages = {27-32},
          doi = {10.18727/0722-6691/5219},
archivePrefix = {arXiv},
       eprint = {2011.12317},
 primaryClass = {astro-ph.IM},
       adsurl = {https://ui.adsabs.harvard.edu/abs/2021Msngr.182...27M}
}

@ARTICLE{2019A&A...631A.155B,
       author = {{Beuzit}, J. -L. and {Vigan}, A. and {Mouillet}, D. and {Dohlen}, K. and {Gratton}, R. and {Boccaletti}, A. and {Sauvage}, J. -F. and {Schmid}, H.~M. and {Langlois}, M. and {Petit}, C. and {Baruffolo}, A. and {Feldt}, M. and {Milli}, J. and {Wahhaj}, Z. and {Abe}, L. and {Anselmi}, U. and {Antichi}, J. and {Barette}, R. and {Baudrand}, J. and {Baudoz}, P. and {Bazzon}, A. and {Bernardi}, P. and {Blanchard}, P. and {Brast}, R. and {Bruno}, P. and {Buey}, T. and {Carbillet}, M. and {Carle}, M. and {Cascone}, E. and {Chapron}, F. and {Charton}, J. and {Chauvin}, G. and {Claudi}, R. and {Costille}, A. and {De Caprio}, V. and {de Boer}, J. and {Delboulb{\'e}}, A. and {Desidera}, S. and {Dominik}, C. and {Downing}, M. and {Dupuis}, O. and {Fabron}, C. and {Fantinel}, D. and {Farisato}, G. and {Feautrier}, P. and {Fedrigo}, E. and {Fusco}, T. and {Gigan}, P. and {Ginski}, C. and {Girard}, J. and {Giro}, E. and {Gisler}, D. and {Gluck}, L. and {Gry}, C. and {Henning}, T. and {Hubin}, N. and {Hugot}, E. and {Incorvaia}, S. and {Jaquet}, M. and {Kasper}, M. and {Lagadec}, E. and {Lagrange}, A. -M. and {Le Coroller}, H. and {Le Mignant}, D. and {Le Ruyet}, B. and {Lessio}, G. and {Lizon}, J. -L. and {Llored}, M. and {Lundin}, L. and {Madec}, F. and {Magnard}, Y. and {Marteaud}, M. and {Martinez}, P. and {Maurel}, D. and {M{\'e}nard}, F. and {Mesa}, D. and {M{\"o}ller-Nilsson}, O. and {Moulin}, T. and {Moutou}, C. and {Orign{\'e}}, A. and {Parisot}, J. and {Pavlov}, A. and {Perret}, D. and {Pragt}, J. and {Puget}, P. and {Rabou}, P. and {Ramos}, J. and {Reess}, J. -M. and {Rigal}, F. and {Rochat}, S. and {Roelfsema}, R. and {Rousset}, G. and {Roux}, A. and {Saisse}, M. and {Salasnich}, B. and {Santambrogio}, E. and {Scuderi}, S. and {Segransan}, D. and {Sevin}, A. and {Siebenmorgen}, R. and {Soenke}, C. and {Stadler}, E. and {Suarez}, M. and {Tiph{\`e}ne}, D. and {Turatto}, M. and {Udry}, S. and {Vakili}, F. and {Waters}, L.~B.~F.~M. and {Weber}, L. and {Wildi}, F. and {Zins}, G. and {Zurlo}, A.},
        title = "{SPHERE: the exoplanet imager for the Very Large Telescope}",
      journal = {\aap},
         year = 2019,
        month = nov,
       volume = {631},
          eid = {A155},
        pages = {A155},
          doi = {10.1051/0004-6361/201935251},
archivePrefix = {arXiv},
       eprint = {1902.04080},
 primaryClass = {astro-ph.IM},
       adsurl = {https://ui.adsabs.harvard.edu/abs/2019A&A...631A.155B}
}

@article {Macintosh12661,
	author = {Macintosh, Bruce and Graham, James R. and Ingraham, Patrick and Konopacky, Quinn and Marois, Christian and Perrin, Marshall and Poyneer, Lisa and Bauman, Brian and Barman, Travis and Burrows, Adam S. and Cardwell, Andrew and Chilcote, Jeffrey and De Rosa, Robert J. and Dillon, Daren and Doyon, Rene and Dunn, Jennifer and Erikson, Darren and Fitzgerald, Michael P. and Gavel, Donald and Goodsell, Stephen and Hartung, Markus and Hibon, Pascale and Kalas, Paul and Larkin, James and Maire, Jerome and Marchis, Franck and Marley, Mark S. and McBride, James and Millar-Blanchaer, Max and Morzinski, Katie and Norton, Andrew and Oppenheimer, B. R. and Palmer, David and Patience, Jennifer and Pueyo, Laurent and Rantakyro, Fredrik and Sadakuni, Naru and Saddlemyer, Leslie and Savransky, Dmitry and Serio, Andrew and Soummer, Remi and Sivaramakrishnan, Anand and Song, Inseok and Thomas, Sandrine and Wallace, J. Kent and Wiktorowicz, Sloane and Wolff, Schuyler},
	title = {First light of the Gemini Planet Imager},
	volume = {111},
	number = {35},
	pages = {12661--12666},
	year = {2014},
	doi = {10.1073/pnas.1304215111},
	publisher = {National Academy of Sciences},
	issn = {0027-8424},
	URL = {https://www.pnas.org/content/111/35/12661},
	eprint = {https://www.pnas.org/content/111/35/12661.full.pdf},
	journal = {Proceedings of the National Academy of Sciences}
}

@INPROCEEDINGS{2020SPIE11447E..1SC,
       author = {{Chilcote}, Jeffrey and {Konopacky}, Quinn and {De Rosa}, Robert J. and {Hamper}, Randall and {Macintosh}, Bruce and {Marois}, Christian and {Perrin}, Marshall D. and {Savransky}, Dmitry and {Soummer}, R{\'e}mi and {V{\'e}ran}, Jean-Pierre and {Agapito}, Guido and {Aleman}, Arlene and {Ammons}, S. Mark and {Bonaglia}, Marco and {Boucher}, Marc-Andre and {Curliss}, Maeve and {Dunn}, Jennifer and {Esposito}, Simone and {Filion}, Guillaume and {Fitzsimmons}, Joeleff and {Kain}, Isabel and {Kerley}, Dan and {Landry}, Jean-Thomas and {Lardiere}, Olivier and {Lemoine-Busserolle}, Marie and {Li}, Duan and {Limbach}, Mary Anne and {Madurowicz}, Alex and {Maire}, Jerome and {N'Diaye}, Mamadou and {Nielsen}, Eric L. and {Poyneer}, Lisa and {Pueyo}, Laurent and {Summey}, Kaitlyn and {Thomas}, Coleman},
        title = "{GPI 2.0: upgrading the Gemini Planet Imager}",
    booktitle = {Ground-based and Airborne Instrumentation for Astronomy VIII},
         year = 2020,
       editor = {{Evans}, Christopher J. and {Bryant}, Julia J. and {Motohara}, Kentaro},
       series = {Society of Photo-Optical Instrumentation Engineers (SPIE) Conference Series},
       volume = {11447},
        month = dec,
          eid = {114471S},
        pages = {114471S},
          doi = {10.1117/12.2562578},
       adsurl = {https://ui.adsabs.harvard.edu/abs/2020SPIE11447E..1SC}
}

@INPROCEEDINGS{2014SPIE.9148E..6CC,
       author = {{Conan}, R. and {Correia}, C.},
        title = "{Object-oriented Matlab adaptive optics toolbox}",
    booktitle = {Adaptive Optics Systems IV},
         year = 2014,
       editor = {{Marchetti}, Enrico and {Close}, Laird M. and {Vran}, Jean-Pierre},
       series = {Society of Photo-Optical Instrumentation Engineers (SPIE) Conference Series},
       volume = {9148},
        month = aug,
          eid = {91486C},
        pages = {91486C},
          doi = {10.1117/12.2054470},
       adsurl = {https://ui.adsabs.harvard.edu/abs/2014SPIE.9148E..6CC}
}

@INPROCEEDINGS{2016SPIE.9909E..7JC,
       author = {{Carbillet}, Marcel and {La Camera}, Andrea and {Folcher}, Jean-Pierre and {Perruchon-Monge}, Ulysse and {Sy}, Adama},
        title = "{The software package CAOS 7.0: enhanced numerical modelling of astronomical adaptive optics systems}",
    booktitle = {Adaptive Optics Systems V},
         year = 2016,
       editor = {{Marchetti}, Enrico and {Close}, Laird M. and {V{\'e}ran}, Jean-Pierre},
       series = {Society of Photo-Optical Instrumentation Engineers (SPIE) Conference Series},
       volume = {9909},
        month = jul,
          eid = {99097J},
        pages = {99097J},
          doi = {10.1117/12.2234280},
       adsurl = {https://ui.adsabs.harvard.edu/abs/2016SPIE.9909E..7JC}
}

@ARTICLE{2010JEOS....5E0055J,
       author = {{Jolissaint}, Laurent},
        title = "{Synthetic modeling of astronomical closed loop adaptive optics}",
      journal = {Journal of the European Optical Society},
         year = "2010",
        month = nov,
       volume = {5},
          eid = {10055},
        pages = {10055},
          doi = {10.2971/jeos.2010.10055},
archivePrefix = {arXiv},
       eprint = {1009.1581},
 primaryClass = {astro-ph.IM},
       adsurl = {https://ui.adsabs.harvard.edu/abs/2010JEOS....5E0055J}
}

@INPROCEEDINGS{2016SPIE.9909E..7FR,
       author = {{Reeves}, Andrew},
        title = "{Soapy: an adaptive optics simulation written purely in Python for rapid concept development}",
    booktitle = {Adaptive Optics Systems V},
         year = 2016,
       editor = {{Marchetti}, Enrico and {Close}, Laird M. and {V{\'e}ran}, Jean-Pierre},
       series = {Society of Photo-Optical Instrumentation Engineers (SPIE) Conference Series},
       volume = {9909},
        month = jul,
          eid = {99097F},
        pages = {99097F},
          doi = {10.1117/12.2232438},
       adsurl = {https://ui.adsabs.harvard.edu/abs/2016SPIE.9909E..7FR}
}

@ARTICLE{2023A&A...674A.207D,
       author = {{Davies}, R. and {Absil}, O. and {Agapito}, G. and {Agudo Berbel}, A. and {Baruffolo}, A. and {Biliotti}, V. and {Black}, M. and {Bonaglia}, M. and {Bonse}, M. and {Briguglio}, R. and {Campana}, P. and {Cao}, Y. and {Carbonaro}, L. and {Cortes}, A. and {Cresci}, G. and {Dallilar}, Y. and {Dannert}, F. and {De Rosa}, R.~J. and {Deysenroth}, M. and {Di Antonio}, I. and {Di Cianno}, A. and {Di Rico}, G. and {Doelman}, D. and {Dolci}, M. and {Dorn}, R. and {Eisenhauer}, F. and {Esposito}, S. and {Fantinel}, D. and {Ferruzzi}, D. and {Feuchtgruber}, H. and {Finger}, G. and {F{\"o}rster Schreiber}, N.~M. and {Gao}, X. and {Gemperlein}, H. and {Genzel}, R. and {Gillessen}, S. and {Ginski}, C. and {Glauser}, A.~M. and {Glindemann}, A. and {Grani}, P. and {Hartl}, M. and {Hayoz}, J. and {Heida}, M. and {Henry}, D. and {Hofmann}, R. and {Huber}, H. and {Kasper}, M. and {Keller}, C. and {Kenworthy}, M. and {Kravchenko}, K. and {Kuntschner}, H. and {Lacour}, S. and {Lightfoot}, J. and {Lunney}, D. and {Lutz}, D. and {Macintosh}, M. and {Mannucci}, F. and {Marsset}, M. and {Modigliani}, A. and {Neeser}, M. and {Orban de Xivry}, G. and {Ott}, T. and {Pallanca}, L. and {Patapis}, P. and {Pearson}, D. and {Pe{\~n}a}, E. and {Percheron}, I. and {Puglisi}, A. and {Quanz}, S.~P. and {Rabien}, S. and {Rau}, C. and {Riccardi}, A. and {Salasnich}, B. and {Schmid}, H. -M. and {Schubert}, J. and {Serra}, B. and {Shimizu}, T. and {Snik}, F. and {Sturm}, E. and {Tacconi}, L. and {Taylor}, W. and {Valentini}, A. and {Waring}, C. and {Wiezorrek}, E. and {Xompero}, M.},
        title = "{The Enhanced Resolution Imager and Spectrograph for the VLT}",
      journal = {\aap},
         year = 2023,
        month = jun,
       volume = {674},
          eid = {A207},
        pages = {A207},
          doi = {10.1051/0004-6361/202346559},
archivePrefix = {arXiv},
       eprint = {2304.02343},
 primaryClass = {astro-ph.IM},
       adsurl = {https://ui.adsabs.harvard.edu/abs/2023A&A...674A.207D}
}

@INPROCEEDINGS{2016SPIE.9909E..0SC,
       author = {{Chin}, Jason C.~Y. and {Wizinowich}, Peter and {Wetherell}, Ed and {Lilley}, Scott and {Cetre}, Sylvain and {Ragland}, Sam and {Medeiros}, Drew and {Tsubota}, Kevin and {Doppmann}, Greg and {Otarola}, Angel and {Wei}, Kai},
        title = "{Keck II laser guide star AO system and performance with the TOPTICA/MPBC laser}",
    booktitle = {Adaptive Optics Systems V},
         year = 2016,
       editor = {{Marchetti}, Enrico and {Close}, Laird M. and {V{\'e}ran}, Jean-Pierre},
       series = {Society of Photo-Optical Instrumentation Engineers (SPIE) Conference Series},
       volume = {9909},
        month = jul,
          eid = {99090S},
        pages = {99090S},
          doi = {10.1117/12.2233138},
       adsurl = {https://ui.adsabs.harvard.edu/abs/2016SPIE.9909E..0SC}
}

@INPROCEEDINGS{1998SPIE.3353..260F,
       author = {{Friedman}, Herbert W. and {Cooke}, Jeff B. and {Danforth}, Pamela M. and {Erbert}, Gaylen V. and {Feldman}, Mark and {Gavel}, Donald T. and {Jenkins}, Sherman L. and {Jones}, Holger E. and {Kanz}, V. Keith and {Kuklo}, Thomas C. and {Newman}, Michael J. and {Pierce}, Edward L. and {Presta}, Robert W. and {Salmon}, J. Thaddeus and {Thompson}, Gary R. and {Wong}, Nan J.},
        title = "{Design and performance of a laser guide star system for the Keck II telescope}",
    booktitle = {Adaptive Optical System Technologies},
         year = 1998,
       editor = {{Bonaccini}, Domenico and {Tyson}, Robert K.},
       series = {Society of Photo-Optical Instrumentation Engineers (SPIE) Conference Series},
       volume = {3353},
        month = sep,
        pages = {260-276},
          doi = {10.1117/12.321681},
       adsurl = {https://ui.adsabs.harvard.edu/abs/1998SPIE.3353..260F}
}

\end{document}